\documentclass[aps,prb,twocolumn,superscriptaddress,floatfix]{revtex4-2}

\usepackage[english]{babel}
\usepackage{amsfonts}
\usepackage{graphicx}
\usepackage{times}
\usepackage[normalem]{ulem} % add this in the preamble

\usepackage{color}
\usepackage{url}
\usepackage{bm,bbm}
\usepackage{graphicx}
\usepackage{xcolor}
\usepackage[colorlinks=true, urlcolor=blue, linkcolor=blue, citecolor=blue, pdftex]{hyperref}
\usepackage[english]{babel}
\usepackage{physics}
\usepackage{slashed}

\usepackage{feynmp-auto}

\begin{document}

\title{Phonon spectral functions of low-density polaron metals}

\author{Luis \surname{Walther}}
\thanks{lwalther@pks.mpg.de}
\affiliation{Max-Planck-Institut f\"{u}r Physik komplexer Systeme, 01187 Dresden, Germany}

\author{Alberto \surname{Nocera}}
\affiliation{Department of Physics and Astronomy, University of British Columbia, Vancouver, BC, Canada, V6T 1Z1}
\affiliation{Quantum Matter Institute, University of British Columbia, Vancouver, BC, Canada, V6T 1Z4}

\author{Mona \surname{Berciu}}
\affiliation{Department of Physics and Astronomy, University of British Columbia, Vancouver, BC, Canada, V6T 1Z1}
\affiliation{Quantum Matter Institute, University of British Columbia, Vancouver, BC, Canada, V6T 1Z4}

\date{\today}
\begin{abstract}
We use the density matrix renormalization group (DMRG) to compute the phonon spectral function of a one-dimensional spinless Holstein model doped with a low but finite carrier concentration, $x \leq 0.15$, as a function of the electron-phonon coupling $\lambda$. To the best of our knowledge, these are the first such results in this regime, complementing extensive prior work at the single-polaron level ($x\to 0$). We find that significant phonon spectral weight is transferred both below, all the way down to $\omega=0$, and above the bare phonon energy $\Omega$, in stark contrast with the Kohn-anomaly phenomenology expected in the Migdal limit, where weight remains centered near $\Omega$ with a kink at $q=2k_F$. No signature of this $2k_F$ kink appears in our results. This behavior is captured qualitatively by the Random Phase Approximation (RPA), and semi-quantitatively, at negligible extra computational cost, by a ``dressed RPA'' scheme in which the electron addition propagator is renormalized using the Momentum Average (MA) approximation for the low-density electron-polaron. By contrast, adding the lowest-order vertex correction to this dressed scheme produces unphysical negative spectral weight, signaling that vertex and propagator dressings must be treated consistently once the propagators are dressed nonperturbatively. Our results provide an efficient approximation for the phonon spectral functions of low-density polaron metals, a regime relevant to weakly doped insulators.
\end{abstract}

%\pacs{ }
\maketitle

\section{Introduction}

Electron-phonon coupling is an unavoidable aspect of the physics of any crystal material, and gives rise to a wide variety of interesting phenomena such as the formation of polarons (a quasiparticle consisting of the electron dressed by its cloud of phonons, which describes the lattice distortion created by the electron in its vicinity) \cite{Landau1933,Landau1948}, conventional BCS superconductivity \cite{BCS1,BCS2} and perhaps also high-temperature phonon-mediated superconductivity through bipolaron condensation~\cite{Zhang2023,Sous2023}, the Peierls transition \cite{Peierls1979}, {\em etc}. All these examples consider primarily the effects of the electron-phonon coupling on the behavior of the electronic subsystem; however, the electron-phonon coupling also affects the behavior of the phonon subsystem. A well-known example of the latter is the Kohn anomaly \cite{Kohn1959} describing the appearance of a ``kink'' in the phonon dispersion at a momentum $q=2k_F$, where $k_F$ is the Fermi momentum of the electrons in the metal. Another example is the appearance of an optical mode at  the Peierls transition in half-filled one-dimensional metals~\cite{Peierls1979,Fomichev2023}. This is the result of the structural instability to dimerization with the period $2a$, driven by a  Kohn anomaly that leads to a complete softening of the phonons with momentum $q=2k_F=\pi/a$.

In metals, the effects of the electron-phonon coupling are well understood in the Migdal limit, defined by a large Fermi energy $E_F$ as compared to the characteristic phonon energy $\Omega$,  $E_F/\Omega \gg 1$, as well as sufficiently weak electron-phonon coupling to avoid polaron formation \cite{Migdal1958,Engelsberg1963}. In this limit one can ignore vertex corrections \cite{Migdal1958}, simplifying the calculation of electron self-energies and phonon polarizations.

In this work, we are interested  in the renormalization of Einstein phonon spectra in metals in the opposite limit $E_F/\Omega \lesssim 1$, which is reached when the carrier concentration $n$ is very small. This limit is relevant for the study of  weakly doped insulators, many of which have been found to have a variety of non-trivial behaviors.

We note that the renormalization of the quasiparticle by the electron-phonon coupling has  been comprehensively studied for a variety of electron-phonon couplings, especially in the limit of a single polaron or single bipolaron (formally, this corresponds to $E_F=0$).  While polaron and bipolaron properties can vary significantly depending on the nature of the electron-phonon coupling, it is by now well established  that  the Migdal theorem fails to describe  their properties accurately, and this is also true for  small carrier concentrations
\cite{Alexandrov2003,Mishchenko2000}. It is therefore to be expected that the Migdal theorem also fails in describing the phonon renormalization in this limit.

The renormalization of the phonon spectra at small carrier concentrations has received much less attention. Numerical results supplemented by analytical perturbative analyses  were presented for the phonon spectral function  in the presence of a single Holstein polaron in Refs.~ \cite{Loos_2006,Barisic2006,Vidmar2010,Jansen2020,rai2026}. This corresponds formally to a finite carrier concentration of $1/N$ for a $N$-site chain, but there is no actual Fermi sea in this single carrier limit: $k_F=0$. While for an Einstein phonon mode of energy $\Omega$, the phonon spectral weight of the undoped chain is non-zero only at the phonon energy, the presence of a polaron was shown to lead to a nontrivial transfer of phonon spectral weight  both below and above $\Omega$~\cite{Loos_2006,Barisic2006,Vidmar2010,Jansen2020,rai2026}. The only other results for the phonon spectral weight that we are aware  of, are for the Holstein model at half-filling~\cite{Weber2015}. Because here the Peierls transition turns the system into an insulator, these results are not relevant for metals.

In this work, we study the phonon spectral weight of a one-dimensional spinless Holstein model at carrier concentrations of up to $x=0.15$, as a function of the electron-phonon coupling. We choose the spinless Holstein model because the spinful system is unstable to bipolaron formation, and a bipolaron liquid is not a normal metal. Of course, bipolaron formation could be prevented by addition of sufficiently strong bare repulsion between carriers; however, it is not \textit{a priori} clear that the resulting metal would be normal, either. This is why the spinless model is the more promising starting point, especially in one dimension.

For this model, we present phonon spectral weights calculated with the density matrix
renormalization group (DMRG). To the best of our knowledge, these are the first such results available in this limit. They confirm that spectral weight is transferred both below and above the bare phonon frequency, consistent with the single polaron results~\cite{Loos_2006,Barisic2006,Vidmar2010,Jansen2020,rai2026}. Even though now $k_F >0$,  there is no sign of a Kohn anomaly at $2k_F$. Considerations based on the lowest order polarization loop allow us to qualitatively understand the main features of these renormalized phonon spectra, including the absence of the Kohn anomaly. We then obtain a better description using the first order polarization loop, but with the electron propagators replaced by the corresponding dressed one-particle propagator calculated with the Momentum Average (MA) approximation for small carrier concentrations~\cite{Berciu2022,Nocera2023}. This improves qualitative agreement everywhere, although for stronger couplings the quantitative difference remains substantial. We then show that including the first vertex correction not only fails to improve agreement, but also produces unphysical (negative) weights; this suggests that crossed diagrams must be summed up to high orders to get quantitative agreement with the DMRG results at stronger couplings.

The paper is organized as follows: Section II describes the model and methods. Section III presents the DMRG results and compares them first with the lowest-order polarization loop (Sec.~III\,A), then with the dressed RPA based on the MA approximation (Sec.~III\,B), and finally with the first vertex correction (Sec.~III\,C). Section IV contains our summary and outlook.

\section{Model and methods}

The 1D spinless Holstein model is described by the Hamiltonian:
\begin{equation}
    {\cal H} = {\hat H}_\mathrm{el}+{\hat H}_\mathrm{ph}+{\hat H}_\mathrm{el-ph}
    \label{e1}
\end{equation}
where ${\hat H}_\mathrm{el}=-t \sum_i (c^\dagger_i c_{i+1} + \mathrm{h.c.})= \sum_k \epsilon(k) c^\dagger_k c_k$ describes nearest-neighbor hopping of the spinless fermions. Here,   $c_i$ ($c_k$) removes a fermion from site $i$ of  (with momentum $k$ from) the chain with $N\rightarrow \infty$ sites and lattice constant $a=1$. The free fermion dispersion is $\epsilon(k)=-2t \cos k$. In the following, we assume that there are $N_f$ fermions in the system, corresponding to the carrier concentration $x= N_f/N$.  The branch of Einstein phonons is described by ${\hat H}_\mathrm{ph}=\Omega \sum_ib_i^\dagger b_i $, where $b_i$ annihilates a phonon from site $i$. Finally, ${\hat H}_\mathrm{el-ph}=g \sum_i c^\dagger_i c_i (b_i + b_i^\dagger) $ is the linear Holstein coupling~\cite{holstein59}, responsible for renormalizing  the properties of both fermions and  bosons. As customary, we use the dimensionless effective coupling $\lambda= g^2/(2t\Omega)$ to characterize the strength of the electron-phonon coupling.

The equations defining the fermion and boson propagators are well established~\cite{Engelsberg1963}. The fermion propagator is:
\begin{equation}
    \label{e2}
    G(k,\omega) = \frac{1}{[G_0(k,\omega)]^{-1} - \Sigma(k,\omega)}
\end{equation}
where the self-energy $\Sigma(k,\omega)$ is defined below, and
\begin{equation}
    \label{e3}
      G_0(k,\omega) = \frac{n_k}{\omega-i\eta- \xi_k}+ \frac{1-n_k}{\omega+i\eta- \xi_k}
\end{equation}
is the free fermion propagator, expressed in terms of the average fermion occupation number $n_k = \langle GS| c_k^\dagger c_k |GS \rangle = \Theta(k_F-|k|)$ with $k_F=x \pi$ and $\Theta(x)$ being the Heaviside function, while $\xi_k = \epsilon(k)-\mu$ is the bare energy shifted by the chemical potential $\mu$.

Similarly, the phonon propagator is:
\begin{equation}
    \label{e4}
    D(q,\omega) = \frac{1}{[D_0(q,\omega)]^{-1} - \Pi(q,\omega)}
\end{equation}
where the polarization $\Pi(q,\omega)$ is defined below, and
\begin{equation}
    \label{e5}
      D_0(q,\omega) = \frac{1}{\omega+i\eta- \Omega}- \frac{1}{\omega-i\eta+ \Omega}
\end{equation}
is the free phonon propagator for an Einstein mode.

For the Holstein coupling, the self-energy and polarization are given by:
\begin{widetext}
\begin{equation}
    \label{6}
\Sigma(k,\omega) = \frac{ig^2}{N} \sum_q \int_{-\infty}^{\infty}\frac{d\omega'}{2\pi}G(k+q,\omega+\omega') \Gamma(k,\omega; q, \omega') D(q,\omega')
\end{equation}
\begin{equation}
    \label{7}
\Pi(q,\omega') = -\frac{ig^2}{N} \sum_k \int_{-\infty}^{\infty}\frac{d\omega}{2\pi}G(k+q,\omega+\omega') G(k,\omega)\Gamma(k,\omega; q, \omega')
\end{equation}
The prefactor of $\Pi(q,\omega')$ does not have the customary factor of 2, because we are considering spinless fermions.
Within the ladder approximation, the vertex is given by the implicit equation:
\begin{equation}
    \label{8}
  \Gamma(k,\omega; q, \omega')  = 1 + \frac{ig^2}{N}\sum_p \int_{-\infty}^{\infty}\frac{d\omega''}{2\pi} G(p+q,\omega''+\omega') \Gamma(p,\omega''; q, \omega') G(p,\omega'') D(k-p,\omega-\omega'')
\end{equation}
\end{widetext}

The quantity of interest is the phonon spectral function:
\begin{equation}
    \label{9}
B(q,\omega) =  -\frac{1}{\pi} \mathrm{Im} D^R(q, \omega)
\end{equation}
where the retarded propagator $D^R(q, \omega)$ is obtained from $D(q,\omega)$ by replacing $\omega-i \eta \rightarrow \omega +i \eta$ everywhere.

\section{Results}

\subsection{DMRG and RPA results}

Equation (\ref{9}) was evaluated directly with DMRG~\cite{WhiteDMRG} using the root-M (we use M here to avoid confusion with $N$, the system size) Krylov correction-vector approach~\cite{re:Nocera2022} as implemented in the DMRG++ software~\cite{re:Alvarez0209}.

For the phonon spectral function, the root-M Krylov method
applies $M$ times the root-$M$ propagator $[\omega-{\mathcal H}+E_{GS,N_e}+i\eta]^{-1/M}$ to the initial vector $d_{x_c}|GS_{N_f}\rangle$, using at each step the standard DMRG Correction Vector Krylov algorithm introduced in Ref.~\cite{re:NoceraPRE2016}. Above, $d_{x_c}=b_{xc}+b^{\dagger}_{xc} -\langle b_{xc}+b^{\dagger}_{xc}\rangle$ is the displacement operator shifted by its ground state expectation value.
As in many previous works, we here employ the center-site approximation, in which the shifted displacement operator is applied at the center site $x_c = N/2$ and the correction vector is locally computed. One then computes $B_{i,x_c}(\omega) = -\frac{1}{\pi}\text{Im}[\langle GS_{N_e}|d_i |\text{CV}_{x_c}\rangle]$ for all sites $i$. These are then used to calculate $B(q,\omega) = \sum_{i=1}^{N}\cos[q(r_i-r_c)]B_{i,x_c}(\omega)$. We have observed that in systems with $N=80$ lattice sites, artifacts from the use of OBC and the center site approximation are minor.

In our DMRG simulations, we used the standard Fock basis representation of the phonon degrees of freedom, and observed that up to $8$ phonon states are sufficient to obtain well converged results.
We therefore did not use the more sophisticated local phonon optimization~\cite{DMRG,Zhang1998,Cheng2012,Brockt2015,Jansen2021,Stolpp2021,DMRGE14,Jansen2022} or the  projected purification approaches~\cite{kohler2021,mardazad2021}. Numerical results were converged with respect to the bond dimension $m$. A maximum $m = 1000$ (and a minimum $m_{\text{min}}=18$) provides convergence with a truncation error  smaller than $10^{-6}$ for the frequency dependent calculations. For the root-$M$ Correction Vector Krylov calculations the choice $M=8$ has shown as the best compromise in terms of moderately large bond dimension required and computational speed.  Finally, we set the Krylov space tridiagonalization error to $\epsilon_{\text{Tridiag}}=10^{-12}$ and $\eta=0.05$ to compute the correction-vectors.

\begin{figure*}
    \centering
    \includegraphics[width=\linewidth]{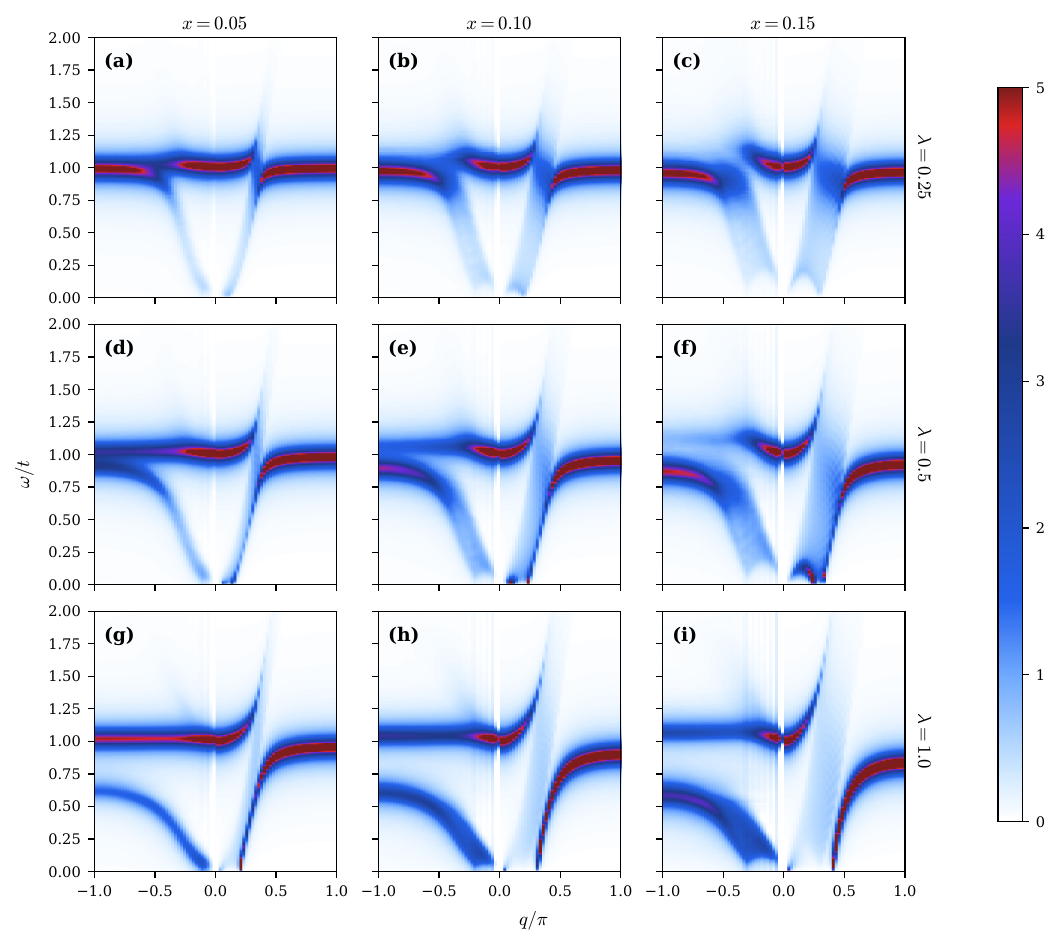}
    \caption{Phonon spectral function $B(q,\omega)$ for $x=0.05$ (left column), $x=0.10$ (middle column) and $x=0.15$ (right column) at $\lambda=0.25$ (top row), $\lambda=0.5$ (middle row) and  $\lambda=1$ (bottom row). In each panel, the left half ($q<0$) shows the DMRG results while the right half ($q>0$) shows results obtained from RPA, {\em i.e.} keeping only the lowest order polarization loop of Eq. (\ref{11}). In all cases $t=1, \Omega=1, \eta=0.05$. }
    \label{fig1}
\end{figure*}

The DMRG results are shown in the left-hand side $(q<0)$ of the panels in Fig. \ref{fig1} at $t=1, \Omega=1, \eta=0.05$ and  for $x=0.05$ (left column), $x=0.10$ (middle column) and $x=0.15$ (right column) at $\lambda=0.25$ (top row), $\lambda=0.5$ (middle row) and  $\lambda=1$ (bottom row).

For comparison, the right-hand side ($q>0$) of all the panels shows the weight obtained from the lowest order contribution to polarization, corresponding to setting $\Gamma \to 1$ and $ G(k,\omega) \to G_0(k,\omega)$ in Eq. (\ref{7}). The expression of the resulting particle-hole polarization loop is evaluated using the residue theorem, to find:
\begin{equation}
    \label{10}
\Pi_0(q,\omega) = \frac{g^2}{N} \sum_k\left[ \frac{n_k (1-n_{k+q})}{z + \xi_k -\xi_{k+q}} -  \frac{n_{k+q} (1-n_{k})}{z^* + \xi_k -\xi_{k+q}} \right]
\end{equation}

We can now obtain the retarded expression by replacing $z^* \to z$, and then simplify it to find:
\begin{equation}
    \label{11}
\Pi^R_0(q,\omega) = \frac{g^2}{N} \sum_k  \frac{n_k - n_{k+q}}{z + \xi_k -\xi_{k+q}}
\end{equation}
This is the well-known random phase approximation (RPA) result~\cite{Mahan}, obeying all the expected symmetries $\Pi^R_0(q,\omega) = \Pi^R_0(-q,\omega) $, $\mathrm{Im } \Pi^R_0(q,\omega)= - \mathrm{Im } \Pi^R_0(q,-\omega)$, etc. The integral over $k$ is carried out numerically.

Within this lowest order RPA approximation, the retarded phonon propagator is given by:
\begin{equation}
    \label{12}
D^R(q,\omega) =\frac{2\Omega}{(\omega+ i \eta)^2 - \Omega^2 - 2 \Omega \Pi_0^R(q,\omega)}
\end{equation}
Its spectral phonon weight, see Eq. (\ref{9}), is the quantity plotted on the right-hand side of the panels in Fig. \ref{fig1}.

The RPA prediction for the phonon spectral function is straightforward to understand physically. It describes processes where a phonon with energy $\omega$ and momentum $q$ is absorbed by a fermion with momentum $k$ inside the Fermi sea, exciting it to a state with momentum $k+q$ above the Fermi surface. Spectral weight is thus possible inside the particle-hole continuum whose lower edge, for a given $q>0$, is $\omega_\mathrm{min}(q) = \min_k[\xi_{k+q}-\xi_k ]= 4t \sin \frac{q}{2} |\sin(k_F - \frac{q}{2})|$, while the upper edge is $\omega_\mathrm{max}(q) =  4t \sin \frac{q}{2} \sin(k_F +\frac{q}{2})$.

The lower edge of the particle-hole continuum goes to zero at $q=0$ and $q=2k_F$, because for a 1D Fermi sea and in the limit of zero energy transfer $\omega \to 0$, only particle-hole pairs with these two momenta can be created. The maximum value for  $\omega_\mathrm{min}(q)$ is reached at $q=k_F$ and equals the bare Fermi energy. This is a direct consequence of the fact that the minimum energy cost for a particle-hole pair with total momentum $k_F$ corresponds to the process where the electron is excited from the center of the Fermi sea to its surface. All these features are seen clearly in the top row of Fig. \ref{fig1}, both in the DMRG and in the RPA results. For any other momentum $q$ we have $\omega_\mathrm{min}(q)>0$  because creating these particle-hole pairs requires a finite minimum energy.  As $q$ increases, this continuum $\omega \in [\omega_\mathrm{min}(q), \omega_\mathrm{max}(q)]$ moves to higher energies. Within RPA, avoided crossings occur when the continuum intersects the bare phonon energy $\Omega$, breaking the latter into a small momentum part that bends upwards, and a large momentum part that bends downwards, beside the continuum that moves through. The weight of the features decreases as they move away from $\Omega$ and couple less to the bare phonons.

The $\lambda=0.25$ panels (top row of Fig. \ref{fig1}) show quantitative agreement between DMRG and RPA, as expected for weak coupling; the one main qualitative  difference is the weight in the continuum as it crosses $\Omega$. However, with increasing $\lambda$ (middle and bottom rows), the differences become significant. RPA predicts additional poles, for instance the large-$q$ part of the spectrum is now a discrete state that starts at $\omega=0$ for a $q > 2k_F$ and disperses upward, flattening for large $q$ to a value close to, but smaller than $\Omega$. This is very different from what DMRG shows at stronger couplings, where the spectrum consists of a strongly renormalized particle-hole continuum that  flattens well below the bare energy $\Omega$, above which there is nearly dispersionless weight at just above $\Omega$, plus some low-$q$ weight dispersing upwards above $\Omega$. Only the latter feature matches with its counterpart from the RPA prediction.

\begin{figure*}
    \centering
    \includegraphics[width=\linewidth]{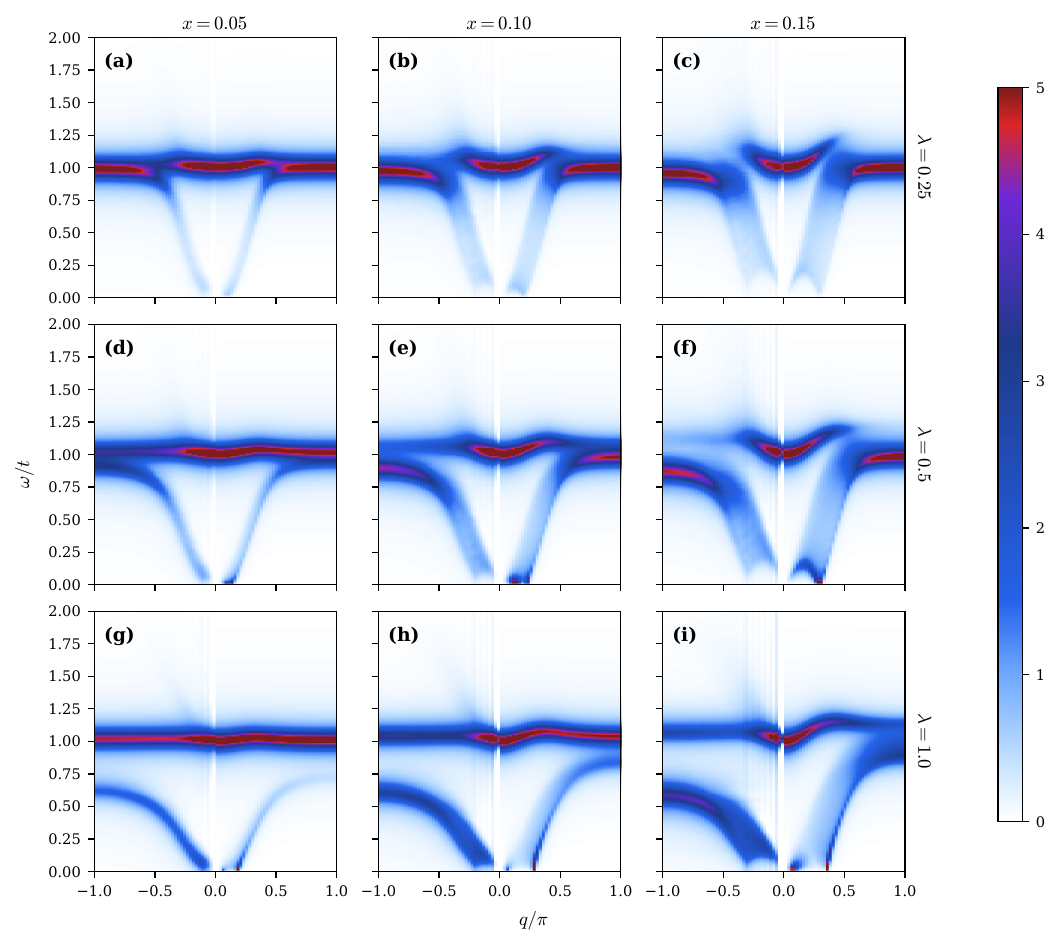}
    \caption{Phonon spectral function $B(q,\omega)$ for $x=0.05$ (left column), $x=0.10$ (middle column) and $x=0.15$ (right column) at $\lambda=0.25$ (top row), $\lambda=0.5$ (middle row) and  $\lambda=1$ (bottom row). In each panel, the left half ($q<0$) shows the DMRG results while the right half ($q>0$) shows results obtained from dressed RPA of Eq. (\ref{15}), see text for more details. In all cases $t=1, \Omega=1, \eta=0.05$. }
    \label{fig2}
\end{figure*}

Before continuing, we emphasize the major differences between these DMRG spectra and the standard Kohn anomaly phenomenology. The DMRG spectra show phonon weight moved from the bare $\Omega$ all the way down to $\omega=0$ as well as above $\Omega$, and there is no signature associated with $2k_F$ at or near $\Omega$. In the standard Kohn anomaly prediction, phonon spectral weight is expected only around $\Omega$, with a kink at $2k_F$. The RPA results confirm that the latter would indeed be recovered for $\Omega/E_F \ll 1$, when the bare phonon dispersion would intersect the particle-hole continuum in the small-$\omega$ limit where the latter is non-zero only near $q=0, 2k_F$.

These DMRG results clearly demonstrate that signatures of the electron-phonon coupling in the phonon renormalization of low-density metals are very different from those in high-density metals.

\subsection{Dressed RPA results}

The significant renormalization of the particle-hole continuum revealed by the DMRG results at medium and strong couplings is not surprising. It is well known that at larger couplings, both the electron above the Fermi sea and the hole left behind inside the Fermi sea, will be dressed with phonon clouds and become an electron-polaron and a hole-polaron, respectively.  Their dispersions are strongly renormalized as $\lambda$ increases, and this must strongly affect the location of the continuum.

The easiest way to test this hypothesis is with a ``dressed RPA'' approximation for the polarization of Eq. (\ref{7}), where we continue to set $\Gamma \to 1$ but use for the electron-addition part of $G(k,\omega)$ the expression derived in Ref. \onlinecite{Berciu2022} using the generalization of the Momentum Average (MA) approximation to low carrier concentrations.  For completeness, we mention that this approximation assumes the metal itself to be ``frozen'' in its mean-field state, which consists of a Fermi sea for the fermions and a uniform distortion for the lattice. The fermion added above the Fermi sea is allowed to build its full polaron cloud, subject to Fermi blockade from the other fermions. Comparisons of the MA single-particle addition spectral weights for the 1D spinless Holstein model with those obtained with DMRG show good agreement for  values of $x, \lambda$ similar to those used here \cite{Nocera2023}.

We therefore replace
\begin{equation}
    \label{13}
    G(k,\omega) = G_A(k, \omega) + G_R(k, \omega)
\end{equation}
where $G_A(k,\omega)$ is the MA$^{(1)}$ Green's function for fermion addition  (non-zero only for $|k|> k_F$), whose expression is identical to that in Ref. \cite{Berciu2022}, and we do not repeat it here. We continue to set the particle-removal propagator  $G_R(k,\omega)$  to its bare value because we do not currently have a better approximation for it.

Because we are primarily interested in understanding the effect of  renormalizing the electron-polaron  energy to see if it improves agreement with DMRG, we further simplify $G_A(k,\omega)$ by keeping only the coherent polaron peak, and discarding the incoherent higher spectral weight (polaron+one-phonon continuum and higher energy features). This additional approximation leads to:
\begin{equation}
    \label{14}
    G(k,\omega) \approx \frac{R(k)}{\omega-i\eta- E_{R}(k)}+ \frac{A(k)}{\omega+i\eta- E_{A}(k)}
\end{equation}
This is similar to Eq. (\ref{e3}), with $R(k) = n_k $ and $E_{R}(k) = \xi_k$  (the bare values), while $A(k) = (1-n_k) Z_{k}$ and  $E_{A}(k) = E_{P}(k) - E_{P}(k_F)$ are set by the electron-polaron's quasiparticle weight $Z_{k}$ and its renormalized electron-polaron energy, measured from its Fermi value. The polaron  quasiparticle weights and the renormalized energies are extracted from fitting Lorentzians of width $\eta$ to the lowest-energy (polaron) peaks of the MA Green's functions.

\begin{figure*}
    \centering
\includegraphics[width=\linewidth]{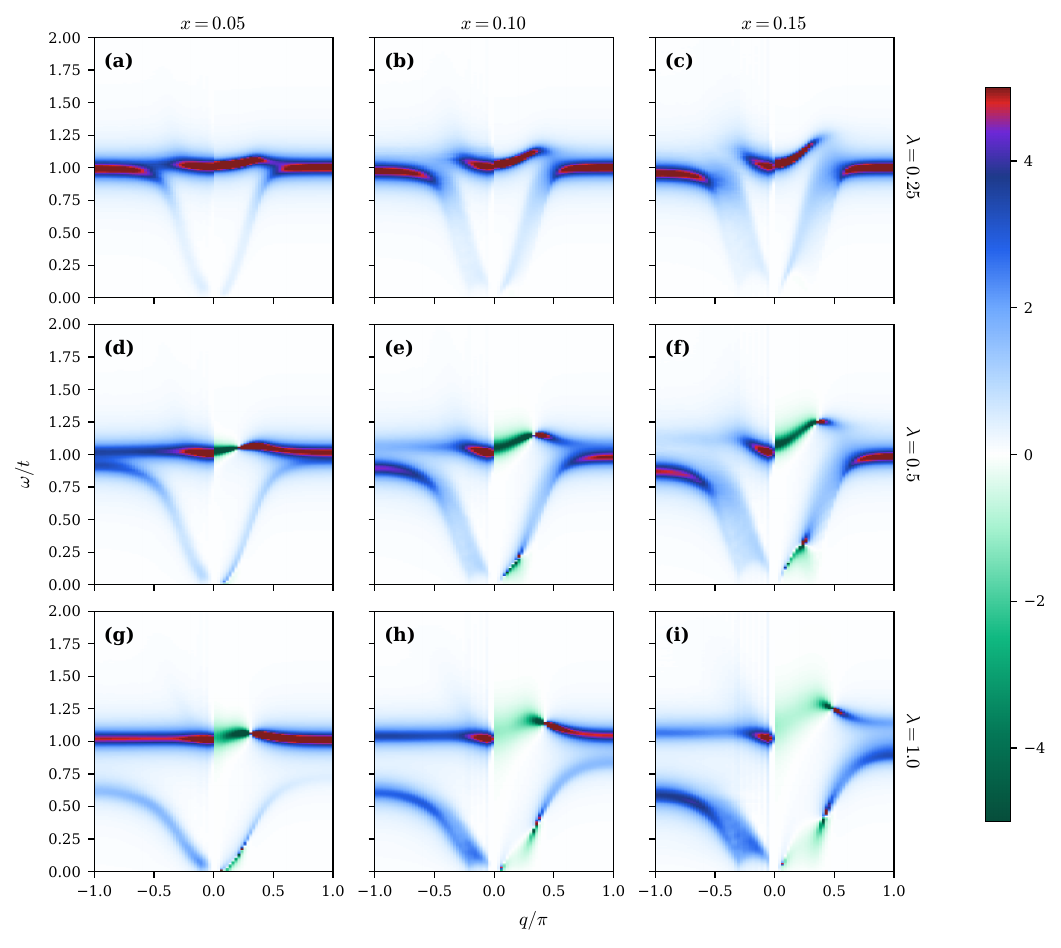}
    \caption{Phonon spectral function $B(q,\omega)$ for $x=0.05$ (left column), $x=0.10$ (middle column) and $x=0.15$ (right column) at $\lambda=0.25$ (top row), $\lambda=0.5$ (middle row) and  $\lambda=1$ (bottom row). In each panel, the left half ($q<0$) shows the DMRG results while the right half ($q>0$) shows results obtained from the polarization that also includes the first vertex correction, see text for more details. In all cases $t=1, \Omega=1, \eta=0.05$.}
    \label{fig3}
\end{figure*}

Using Eq. (\ref{14}) in the polarization integral Eq. (\ref{7}) while continuing to set $\Gamma\to 1$, leads to the dressed RPA result:
\begin{equation}
\begin{aligned}
    \tilde{\Pi}_0({q}, \omega)=\frac{g^2}{N}\sum_{{k}}&\Bigg[\frac{A({k + q})R({k})}{z - E_{A}(k+q) +E_{R}(k)}\\
    +&\frac{A(k)R(k+q)}{ - z^*- E_{A}(k) + E_{R}( k+q)}\Bigg]
\end{aligned}
\label{15}
\end{equation}

%\begin{align}
 %   \tilde{\Pi}_0({q}, \omega)=\frac{g^2}{N}\sum_{{k}}&\Bigg[\frac{A({k + q})R({k})}{z - E_{A}(k+q) +E_{R}(k)}\nonumber\\
  %  +&\frac{A(k)R(k+q)}{ - z^*- E_{A}(k) + E_{R}( k+q)}\Bigg]\label{15}\raisetag{515.5\baselineskip}
%\end{align}
The retarded expression is obtained by replacing $z^* \to z$, and the retarded phonon propagator is then calculated by using this retarded polarization in Eq. (\ref{12}). The corresponding phonon spectral weights are shown on the right-hand side ($q>0$) of the panels in Figure \ref{fig2}, while the left-hand side displays again the DMRG results, for ease of comparison. All parameters are the same as in Fig. \ref{fig1}.

This ``dressed RPA'' shows qualitative  improvement over the lowest-order RPA. Especially at stronger couplings, the particle-hole continuum is strongly renormalized and flattens out below the fairly dispersionless weight located just above $\Omega$. The agreement is quantitatively reasonable for $x=0.05$ but worsens with increasing $x$, {\em e.g.} for $x=0.15, \lambda=1$, at the Brillouin-zone edge the continuum  is spread over $\sim [0.8,1]$ instead of the DMRG location $ \sim [0.5,0.75]$.

Based on these results, we conclude that the ``dressed RPA'' approximation works reasonably well for predicting the renormalized phonon spectra, especially considering its simplicity and computational efficiency. Moreover, the MA approximation is known to improve quantitatively in higher dimensions \cite{Goodvin2006}, so this method can be easily extended to study low-density polaron liquids in higher dimension with improved  accuracy. To our knowledge, no such results have been published yet. Finally, the derivation of a better approximation for the removal part of the Green's function in the limit of low carrier concentration should further improve the agreement.

\subsection{First vertex correction}
\label{sec:vertex}

We also investigated the possibility of further improving the agreement between DMRG and analytical methods by including the lowest vertex correction, {\em i.e.} instead of setting $\Gamma\to 1$, to use either:
\par
\begin{widetext}
\begin{equation}
    \label{16}
  \Gamma(k,\omega; q, \omega')  = 1 + \frac{ig^2}{N}\sum_p \int_{-\infty}^{\infty}\frac{d\omega''}{2\pi} G_0(p+q,\omega''+\omega')  G_0(p,\omega'') D_0(k-p,\omega-\omega'')
\end{equation}
or
\begin{equation}
    \label{17}
  \Gamma(k,\omega; q, \omega')  = 1 + \frac{ig^2}{N}\sum_p \int_{-\infty}^{\infty}\frac{d\omega''}{2\pi} G(p+q,\omega''+\omega')  G(p,\omega'') D_0(k-p,\omega-\omega''),
\end{equation}
\end{widetext}
where the simplified MA expression from Eq. (\ref{14}) is used for $G(k,\omega)$ in Eq. (\ref{17}). The resulting vertex can be inserted in Eq. (\ref{7})  together with the bare propagators $G_0(k,\omega)$, or with the $G(k,\omega)$ of Eq. (\ref{14}), respectively.

Within this approximation two energy integrals appear (one in $\Gamma$, one in $\Pi$), which both can  be carried out analytically, resulting in a rather lengthy formula for the polarization. For completeness, we give its expression in the Appendix for the second option of using the MA$^{(1)}$ $G(k,\omega)$. The expression corresponding to $G_0(k,\omega)$ can be obtained by replacing  $A(k) \to 1- n_k$ and using the bare energy $\epsilon(k)$ instead of the electron-polaron's energy $E_{A}(k)$. The final two momentum integrals are carried out numerically; this is an efficient task because each term has at least one $R(k)$-type prefactor, which has non-zero support only inside the small Fermi sea.

The results obtained using the dressed $G(k,\omega)$  option are shown in Figure \ref{fig3}. They are un-physical, predicting negative phonon spectral weight at low momenta for all $x,\lambda$ (the green regions, harder to see for the smaller $\lambda$ because the negative values are smaller in magnitude than for larger $\lambda$). Such negative weight also appears when we use the $G_0(k,\omega) $ propagators, hinting at the fact that this is an issue with the approximation. Given that the negativity of spectral weights increases with increasing $\lambda$, we believe that this signals inconsistency in keeping powers of $\lambda \sim g^2$ contributions. For instance, when we use the bare $G_0(k, \omega)$, the vertex $\Gamma$ contains contributions up to ${\cal O}(g^2)$, however using $G_0(k,\omega)$ in $\Pi$ misses the similar contributions of order ${\cal O}(g^2)$ coming from the fermion propagators' renormalization.

Counting powers is more complicated when using the dressed $G(k,\omega)$ expression because these fermion propagators contain contributions up to arbitrarily high powers of $g^2$. In principle so does the vertex correction through its dependence on $G(k,\omega)$, however the phonon contribution $D_0$ is still the bare value. A possible future approach is to use iterations to solve Eqs. (\ref{7}) and (\ref{8}) until self-consistency is reached, to see if this re-establishes strictly positive spectral weights. 

We note that a recent study by Rai and Pandey~\cite{rai2026} also examined the effect of vertex corrections on the Holstein polaron phonon spectral function, using a weak-coupling, Ward-identity-consistent diagrammatic scheme. That work addresses the single-polaron limit ($x\to 0$) in the antiadiabatic regime, and finds that the vertex correction there {\em enhances} the polaronic spectral weight, partially compensating the suppression coming from the electron self-energy; this is a well-behaved, physical result at weak coupling. This is not inconsistent with our findings: at finite carrier concentration, we combine the lowest-order vertex correction with electron propagators that are already dressed nonperturbatively via the MA approximation, which, as discussed above, mixes contributions of different order in $g^2$ in an uncontrolled way and is the most likely origin of the unphysical negative weights we obtain. The comparison suggests that a consistent, Ward-identity-respecting treatment of the vertex, of the type used in Ref.~\cite{rai2026}, will likely be needed to extend our dressed RPA results to a fully controlled diagrammatic scheme at finite $k_F$.

\section{Conclusions}

In this work we used DMRG to compute the phonon spectral function of a one-dimensional spinless Holstein model  at low but finite carrier concentrations $x \leq 0.15$. We find that the renormalization of the phonon spectrum in this regime is qualitatively distinct from the Migdal-limit phenomenology captured by the traditional Kohn-anomaly picture: rather than remaining sharply peaked near the bare phonon energy $\Omega$ with a kink at $q=2k_F$, spectral weight is transferred continuously away from $\Omega$, extending all the way down to $\omega=0$ as well as above $\Omega$, with no discernible signature at $2k_F$. This shows that the standard Migdal-Eliashberg intuition for phonon renormalization does not carry over to the low-density limit, even at the qualitative level.

We show that the main features of this renormalized spectrum are captured, at essentially no extra computational cost, by a simple one-loop (RPA-like) approximation for the polarization, provided the bare electron propagators entering the loop are replaced by propagators ``dressed'' via the Momentum Average (MA) approximation for the low-density electron-polaron. This dressed RPA scheme reproduces the DMRG results semi-quantitatively across the couplings and concentrations studied, with the residual disagreement growing with $x$ and $\lambda$. By contrast, we found that including the lowest-order vertex correction, whether evaluated with bare or dressed propagators, produces unphysical negative phonon spectral weight, indicating an inconsistent truncation of the diagrammatic series. As discussed in Sec.~\ref{sec:vertex}, this is consistent with the different, weak-coupling and Ward-identity-consistent treatment used in Ref.~\cite{rai2026} for the single-polaron limit. We did not find this vertex correction to offer any improvement over the dressed RPA, even in the regions where its weight remains positive.

Taken together, these results offer a practical guideline for efficiently approximating phonon spectral functions in this low-density regime: strong electron-phonon coupling effects seem to be reasonably  well captured by using dressed (polaronic) propagators in the lowest-order polarization loop, while the vertex itself can be safely set to its bare value, $\Gamma=1$, just as in the Migdal limit. We do not yet have a physical explanation for why this scheme works as well as it does in this regime, nor a criterion for when it should start to fail not just quantitatively but also qualitatively.

Looking forward, because the MA approximation is known to become even more  accurate in higher dimensions~\cite{Goodvin2006}, the dressed RPA scheme employed here should extend naturally to low-density polaron liquids in two and three dimensions, where no comparable results are yet available. Further improvement will come from a better treatment of the hole-removal part of the fermion propagator, which we have left at its bare value in this work, potentially together with a fully self-consistent, Ward-identity-respecting treatment of the vertex. We leave both directions for future study.

\begin{acknowledgments}
 This project was supported by the Max Planck--UBC--UTokyo Center for Quantum Materials and by the Natural Sciences and Engineering Research Council of Canada (M.B. and A. N.). % TODO: add L.W. fellowship/funding acknowledgment, if applicable
\end{acknowledgments}

\appendix

\section{Polarization including the first vertex correction }

The expression we obtain for the polarization when including the first vertex correction using the MA approximation of Eq. (\ref{14}) for the fermion propagators is:
\begin{equation}
    \label{ap1}
    \Pi(q, \omega) = \tilde{\Pi}_0(q,\omega) + \tilde{\Pi}_1(q,\omega)
\end{equation}
where $ \tilde{\Pi}_0(q,\omega)$ is the ``dressed RPA'' expression from Eq. (\ref{15}), while:
\begin{widetext}
    \begin{equation}
        \begin{split}
            \tilde{\Pi}_1({q}, \omega)=\frac{2g^4}{N^2}\sum_{k,p} \Bigg[& - \frac{R({k})R({k}+{q})A({p})R({p}+{q})}{\Omega + E_A({p}) -E_R({k}) - 3 i \eta}\Bigg(\frac{1}{z_2^* +E_A({p})-E_R({p}+{q})}\cdot\frac{1}{z_3^*+ E_A({p})-E_R({k}+{q})+\Omega}\\
            &\phantom{aaaaaaaaaaaaaaaaaaaaaa}+\frac{1}{- z_2 +E_A({p})- E_R({p}+{q})}\cdot\frac{1}{- z_3 +E_A({p}) -E_R({k}+{q}) + \Omega}\Bigg)\\
            &\\
            &+\frac{R({k}) R({k}+{q}) A({p})A({p}+{q})}{\Omega + E_A({p}) -E_R({k}) - 3 i \eta}\cdot \frac{1}{\Omega + E_A({p}+{q}) -E_R({k+q}) - 3 i \eta}\\ &\phantom{aaaaaaaaaaaaaaaaa}\times\Bigg(\frac{1}{- z_3 +E_A({p}+{q}) + \Omega -E_R({k})}+\frac{1}{z_3^*+ \Omega + E_A({p}+{q}) -E_R({k})}\Bigg)\\
            &\\
            &+\Bigg(\frac{R({k}) A({k}+{q})  R({p})A({p}+{q})}{z_2^* +E_A({p}+{q}) -E_R({p})}\cdot\frac{1}{z_2^* +E_A({k}+{q}) -E_R({k})}\cdot\frac{1}{z_3^* + \Omega +E_A({p}+{q}) -E_R({k})}\\
            &+\frac{R({k}) A({k}+{q})  R({p})A({p}+{q})}{- z_2 +E_A({p}+{q}) -E_R({p})}\cdot\frac{1}{- z_2 +E_A({k}+{q}) -E_R({k})}\cdot\frac{1}{- z_3 + \Omega+ E_A({p}+{q}) -E_R({k})}\Bigg)\\
            &\\
            &+\frac{R({k})A({k}+{q})  A({p})  R({p}+{q})}{\Omega + E_A({p}) -E_R({k}) - 3 i \eta}\Bigg(\frac{1}{- z_2+E_A({k}+{q}) -E_R({k})}\cdot\frac{1}{z_2^* + E_A({p})-E_R({p}+{q})}\\
            &\phantom{aaaaaaaaaaaaaaaaaaaaaaaaaaa}+\frac{1}{z_2^* +E_A({k}+{q}) -E_R({k})}+\frac{1}{- z_2 +E_A({p}) -E_R({p}+{q})}\Bigg)\\
            &\\
            &-\frac{R({k}) A({k}+{q}) A({p}) A({p}+{q})}{\Omega +E_A({p}) -E_R({k}) - 3 i \eta}\Bigg(\frac{1}{z_2^* +E_A({k}+{q}) -E_R({k})}\cdot\frac{1}{z_3^*+E_A({p}+{q}) -E_R({k}) + \Omega}\\
            &\phantom{aaaaaaaaaaaaaaaaaaaaaaaa}+\frac{1}{-z_2 + E_A({k}+{q}) -E_R({k})}\cdot\frac{1}{- z_3+E_A({p}+{q}) -E_R({k}) + \Omega}\Bigg)\Bigg]
        \end{split}
    \end{equation}
\end{widetext}
where $z_2 = \omega + 2 i\eta$ and $ z_3= \omega + 3 i \eta$. We note that in the limit $\eta \to 0$ there is no difference between $\omega + i\eta$ and $\omega +2i\eta $, but because we use a finite $\eta=0.05$, we keep track of the correct location of the poles and use it in all numerical results including the first vertex correction. The retarded expression is obtained by replacing $z_2^* \to z_2, z_3^*\to z_3.$

\bibliography{article}
\end{document}